\documentclass[conference]{IEEEtran}

\newcommand{\Prob}{I\!\!P}
\usepackage{graphicx}

\usepackage{amsfonts}
\usepackage{amssymb}
\usepackage{amsmath}
\usepackage{float}
\usepackage[hang,small,bf]{caption}                   % Default: 10pt Times Roman fonts
\DeclareGraphicsExtensions{.pdf,.png,.jpg}
\def\endpf{\hfill$\diamond$}

\title{Cross-Layer Design for Green Power Control }

\author {Vineeth S Varma$^{1,2}$, Samson Lasaulce$^2$, Yezekael Hayel$^3$, Salah Eddine Elayoubi$^1$ and Merouane Debbah$^4$ \fontsize{9}{8} \\\\ \begin{minipage}{131pt} \centering $^{1}$Orange Labs\\92130 Issy Les Moulineaux\\France\\salaheddine.elayoubi@orange.com
\end{minipage}
\begin{minipage}{30pt}
\end{minipage}
\begin{minipage}{130pt}\centering  $^{2}$LSS, SUPELEC\\ 91192 Gif sur Yvette\\ France\\ Vineeth.varma@lss.supelec.fr\\
Samson.lasaulce@lss.supelec.fr\\

\end{minipage}
\begin{minipage}{30pt}  \space
\end{minipage}
\begin{minipage}{120pt}\centering  $^{3}$University of Avignon, \\84911 Avignon\\France\\yezekael.hayel@univ-avignon.fr
\end{minipage}\begin{minipage}{120pt}\centering  $^{4}$Alcatel Lucent Chair\\SUPELEC\\ 91192 Gif sur Yvette\\ France\\
Merouane.debbah@supelec.fr
\end{minipage}}

\begin{document}
\maketitle

\begin{abstract}

In this work, we propose a new energy efficiency metric which allows one to optimize the performance of a wireless system through a novel power control mechanism. The proposed metric possesses two important features. First, it considers the whole power of the terminal and not just the radiated power. Second, it can account for the limited buffer memory of transmitters which store arriving packets as a queue and transmit them with a success rate that is determined by the transmit power and channel conditions. Remarkably, this metric is shown to have attractive properties such as quasi-concavity with respect to the transmit power and a unique maximum, allowing to derive an optimal power control scheme. Based on analytical and numerical results, the influence of the packet arrival rate, the size of the queue, and the constraints in terms of quality of service are studied. Simulations show that the proposed cross-layer approach of power control may lead to significant gains in terms of transmit power compared to a physical layer approach of green communications.
\end{abstract}

\section{Introduction}

For a long time, the problem of energy mainly concerned autonomous, embarked, or mobile communication terminals. Nowadays, with the existence of large networks involving both fixed and nomadic terminals, the energy consumed by the fixed infrastructure has also become a central issue for communications engineers \cite{salah}. The present work precisely falls into this framework. More specifically, our goal is to provide insights to researchers and engineers on how to devise power control schemes in green wireless networks. Among pioneering works on energy-efficient power control we find the paper by Goodman et al \cite{goodman}. Therein, the authors define the energy-efficiency of a communication as the ratio of the net data rate (called goodput) to the radiated power; the corresponding quantity is a measure of the average number
of bits successfully received per joule consumed at the transmitter. This metric has been used in many works. For example, in \cite{meshkatiCDMA06} it is applied to the problem of distributed power allocation in multi-carrier CDMA (code division multiple access systems) systems, in \cite{Mesh09} it is used to model the users delay requirements in energy-efficient systems. In \cite{veronica2} it is re-interpreted as a capacity per unit cost measure in MIMO (multiple input multiple output) systems and in \cite{valuetools}, it is used for subcarrier assignment in distributed OFDMA (orthogonal frequency division multiple access) multicellular networks. Although fully relevant, the performance metric introduced in \cite{goodman} has left several issues unexplored which has motivated the work reported here.

First, in the definition of energy-efficiency given in works like \cite{rodriguez},\cite{goodman} or \cite{veronica2}, the transmission cost corresponds to the radiated power that is, the power of the radio-frequency signals; this is very useful in situations where electromagnetic pollution has to be cut down. However, more generally, the consumption of the whole device matters (e.g., because of the power amplifier consumption). Second, in \cite{goodman}, the packets are lost due to bad channel conditions while, here, we propose to take into account the losses induced by the finite size of the queue at the transmitter (which can model limited memory or a certain delay constraint). Third, in \cite{goodman} and related references, energy-efficiency can be maximized while having a bad quality of service (QoS) e.g., in terms of packet loss or equivalently in terms of goodput. In this paper, we show that these three issues can be, in fact, dealt with quite easily. The analysis is, however, more complicated than some analysis like the one conducted in \cite{Mesh09} where the delay constraint is translated into a constraint on the minimum signal-to-noise ratio (SNR). This is due to a double effect, resulting from integrating a queueing model (justifying the term ``cross-layer design") and considering the whole terminal power (instead of the radiated power). The queuing model is used in the spirit of \cite{Kim00} where a queuing model is used to reach a certain QoS in CDMA systems with multiple classes of traffic, but without energy considerations. Another cross-layer queuing model has been proposed in \cite{AltmanCons07} but considering Shannon capacity under power constraints and not energy-efficiency.

At this point, it is important to note that this work focuses on point-to-point communications, which may be surprising since power control is the problem of interest. There are at least two strong reasons for this choice. First, the single-user case captures the main effects we want to emphasize and allows us to describe the proposed approach in a clear manner. Second, as advocated by the existing works on power control (see e.g., \cite{goodman} and related works), once the single-user case is fixed, the multiuser case is tractable provided some conditions are met. One of these conditions is that the performance metric has to possess some desirable properties (quasi-concavity, that is shown to be verified for the proposed metric, is one of them) and reasonably complex multiuser channel models are considered (the multiple access channel is one of them).

This paper is structured in the following format. In Section II, we present the system model and define the proposed performance metric. In Section III, we conduct an analytical study of the performance metric while Section IV provides many numerical results to sustain the proposed approach. Finally, we conclude the paper and some possible extensions to this work are provided.

\section{System model}

We consider a buffer of size $K$ at the transmitter. The packets arrival follows a Bernoulli process with probability $q$, i.e., at each time slot $t$ (time is slotted) a new packet arrives in the queue with probability $q$ (this corresponds to classical ON/OFF sources). All packets are assumed to be of the same size $S$. The throughput on the radio interface equals to $R$ (bit/s) and depends on several parameters such as the modulation and coding scheme. We consider that the transmitter is always active, meaning that it always transmits its packet while the buffer is not empty. Each packet transmitted on the channel is received without any errors with a probability which depends on the quality of the channel and transmission power. We denote the transmission power by $p$ and we have $f(p)$ as the success probability of transmission of the packet on the channel. $f(p)$ is just assumed to be a sigmoidal function in our derivations, in practice, good approximations for the success rate function are $f(p)=\exp(-(2^{\frac{R}{R_0}}-1)\frac{\sigma^2}{p})$, for an unknown channel, and  $f(p)=Q(K\frac{R}{R_0}-K\log[1+\frac{hh^* p}{\sigma^2}])$, with $Q$ being the ``Q" function and $K$ a constant, for a known channel. The channel fading due to path loss is not treated separately but is integrated into the average power of noise $\sigma^2$. The success probability depends in fact on the $\mathrm{SNR} = \frac{p}{\sigma^2}$. However, based on the block fading channel assumption, we make a slight abuse of notations by using the notation $f(p)$ instead of $f(\mathrm{SNR})$. In some places in this paper, we even remove the variable $p$ for the sake of clarity and use the notation $f$. We denote by $Q_t$ the size of the queue at the transmitter at time slot $t$. The size of the queue $Q_t$ is a Markov process on the state space $Q=\{0,\ldots,K\}$. We have the following transition probabilities $\forall i,j \in Q$, $P_{i,j}:=\Prob(Q(t+1)=i|Q(t)=j)$ given by:
\begin{itemize}
  \item $P_{0,0}=1-q+qf$,
  \item $P_{K,K}=(1-q)(1-f)+q$,
  \item for any state $i \in \{0,\ldots,K-1\}$, $P_{i,i+1}=q(1-f)$,
  \item for any state $i \in \{1,\ldots,K\}$, $P_{i,i-1}=(1-q)f$,
  \item for any state $i\in \{1,\ldots,K-1\}$, $P_{i,i}=(1-q)(1-f)+fq$.
\end{itemize}

A new packet is lost if the queue is full when it comes in and the transmission of the packet currently on the radio interface failed on the same time slot. Indeed, we consider that a packet is in service (occupying the radio interface) until it is transmitted successfully. Thus, a packet in service blocks the queue during $\frac{1}{f(p)}$ time slots on the average. We assume that an arrival of a packet in the queue and a departure (successful transmission) at the same time slot can occur.

Given the transition probabilities above, the stationary probability of each state is given by (see e.g., \cite{wolff}):
\begin{equation}
\forall s \in S, \quad \Pi_s=\frac{\rho^s}{1+\rho+\ldots+\rho^K},
\end{equation}
with 
\begin{equation}
\rho=\frac{q(1-f)}{(1-q) f }.
\end{equation}
When a packet arrives and finds the buffer full (meaning that the packet currently on the radio interface is not transmitted successfully), it is blocked and this event is considered as a packet loss. The queue is full in the stationary regime with probability $\Pi_K$~:
\begin{equation}
\Pi_K=\frac{\rho^K}{1+\rho+\ldots+\rho^K}=\frac{\rho^K (\rho-1)}{\rho^{K+1}-1}.
\label{eq:pik}
\end{equation}

\subsection{Proposed performance metric}

In order to evaluate the performance of this system, we first determine the expression for the packet loss probability. A packet is lost (blocked) only if a new packet arrives when the queue is full and, on the same time slot, transmission of the packet on the radio interface failed. Note that these two events are independent because the event of ``transmit or not'' for the current packet on the radio interface, does not impact the current size of the queue, but only the one for the next time slot. This amounts to considering that a packet coming at time slot $t$, is rejected at the end of time slot $t$, the packet of the radio interface having not been successfully transmitted. We consider the stationary regime of the queue and then, the fraction of lost packets, $\Phi$, can be expressed as follows:
\begin{equation}
\Phi(p) =  [ 1 - f(p) ] \Pi_K(p).
\label{eq:phi}
\end{equation}
Thus the average data transmission rate is $q[1-\Phi(p)]R$. Now, let us consider the cost of transmitting. For each packet successfully transmitted, there have been $\frac{1}{f(p)}$ attempts on an average \cite{goodman}. For each time slot, irrespective of whether transmissions occur, we assume that the transmitter consumes energy. A simple model which allows one to relate the radiated power to the total device consumed power is provided in \cite {richter} is given by $P_{\mathrm{device}} = a p + b$, where $a \geq 0, b \geq 0$ are some parameters; $b $ precisely represents the consumed power when the transmit power is zero. The average power consumption is in our case $b + \frac{pq(1-\Phi)}{f(p)}$ (we assume without loss of generality that $a=1$). We are now able to define the energy-efficiency metric $\eta(p)$ as the ratio between the average net data transmission rate and the average power consumption, which gives:
\begin{equation}
\label{eq:eta}
\eta(p) =  \frac{q[1-\Phi(p)]R}{b + \frac{pq[1-\Phi(p)]}{f(p)}}.
\end{equation}
The above expression shows that the cross-layer design approach of power control is fully relevant when the transmitter has a cost which is independent of the radiated power; otherwise (when $b=0$), one falls into the original framework of \cite{goodman}.

\subsection {Constraints on QoS and maximum power}

As already mentioned in Section I, of the recurrent problems with most works using the performance metric introduced in \cite{goodman} is that energy-efficiency can be maximized at a power level which does not guarantee a minimum QoS. This is why we also consider a constraint when maximizing (\ref{eq:eta}): we assume $\Pi_K [1-f(p)]$ has to be less than $\epsilon$ where $\epsilon$ is the upper bound on the packet loss. For example, in cellular systems, typical values for $\epsilon$ are $0.1$ or $0.01$, based on the system requirements. Adding this constraint restricts the range of power usable by the transmitter by adding a lower bound on the power. This lower bound depends on the entry probability $q$ and on the size of the queue $K$. An upper bound on the usable transmit power $P_{\mathrm{max}}$ can also be added to model realistic situations when there is a limitation on the maximum power that can be utilized.

\section{Analytical results}

Having defined the energy efficiency function, we will now examine its properties.

\subsection{The impact of the packet entry probability}

First let us study the special case when $q = 1$: Here we have, $\lim_{q \to 1} \Pi_K=1$, then $\Phi=1-f(p)$ and we have a simplified expression of the energy efficiency function $\eta=\frac{Rf(p)}{b + p}$. This energy efficiency is a more general form of the metric introduced in \cite{goodman}. This particular case is in fact identical to the situation when the system is modeled with a purely physical layer approach. As the queue is always full, transmission always takes place and so the energy efficiency of the entire system is just the energy efficiency of transmission.

\newtheorem{propposition}{Proposition}

The next part of this section examines $\eta$ as $q$ decreases. Logically, as $q$ decreases, the average duration when the buffer is empty increases causing a wasted consumption of the fixed power during which no data is transmitted. Here, we have a proposition that formulates and proves this reasoning mathematically.

\begin{propposition}
The energy efficiency function is an increasing function of $q$, i.e., $\frac{\mathrm{d}\eta}{\mathrm{d}q} > 0$.
\end{propposition}

{\bf Proof :} $\eta=\frac{1}{\frac{b}{(1-\Phi)q} + \frac{p}{f(p)}}$. If $\frac{\mathrm{d}\Phi}{\mathrm{d}q} < \frac{1-\Phi}{q}$, then $\mathrm{d}\frac{(1-\Phi)q}{\mathrm{d}q} > 0 $ and from this, it follows that $\frac{\mathrm{d}\eta}{\mathrm{d}q} > 0$. \\

To prove this, we first calculate $\frac{\mathrm{d}\rho}{\mathrm{d}q} =  \frac{1-f(p)}{f(p)} \frac{-1}{(1-q)^2}$. Now let us consider $\frac{\mathrm{d}\Phi}{\mathrm{d}q}=-\Phi^2\frac{\mathrm{d}(\Phi^{-1})}{\mathrm{d}q}$.
The term $\Phi^{-1}=1+\frac{1}{\rho} +..+\frac{1}{\rho^K}$ and so we have $\frac{\mathrm{d}\Phi^{-1}}{\mathrm{d}q}=(\frac{1}{\rho^2} +..+\frac{K}{\rho^{K+1}} )\frac{1-f(p)}{f(p)} \frac{-1}{(1-q)^2}$. Simplifying and using inequalities we have $\frac{\mathrm{d}\Phi}{\mathrm{d}q} \leq \frac{1-\Phi}{q}$.\\
\endpf

\subsection {The limiting case of infinite queue size}
Consider the extreme case of an infinite queue, i.e., $K \to \infty$.
\begin{itemize}
\item Case 1: $f(p) < q$; i.e., $\rho>1$.
We have that $\lim_{K \to \infty} \Pi_K = \frac{\rho-1}{\rho} $ and a simplification yields $\Phi = \frac{1-f(p)}{q}$. Thus the energy efficiency becomes $\eta= \frac{Rf(p)}{b + p}$. These expressions make sense as in the steady state, due to a higher probability of entrance than exit, the queue size blows up and there are always packets to transmit.

\item Case 2: $f(p) \geq q$; i.e., $\rho \leq 1$.
If $f(p)=q$, then $\Pi_K = \frac{1}{K} $ and $\lim_{K \to \infty} \Pi_K =0$. For $f(p)>q$, we have also that $\lim_{K \to \infty} \Pi_K =0 $ and then simplification yields $\Phi = 0$. Thus the energy efficiency becomes $\eta= \frac{R}{\frac{b}{q} + \frac{p}{f(p)}}$. These expressions also make sense as in the steady state due to a higher probability of exit, the buffer is never full and there is no packet loss.
\end{itemize}

\subsection {Optimizing the energy efficiency}

In this paragraph, we prove that there exists a unique power where the energy efficiency function is maximized when the transmission rate is a sigmoidal or "S"-shaped function of $p$. In \cite{rodriguez}, it was shown that having a sigmoidal success rate $f(p)$ implies quasi-concavity and a unique maximum for $\frac{f(p)}{p}$. This assumption was shown to be highly relevant from a practical viewpoint in \cite{goodman} as well as from an information theoretical viewpoint in \cite{veronica2}.

\newtheorem{theory}{Theorem}

\begin{theory}
The energy efficiency function $\eta$ is quasi-concave with respective to $p$ and has a unique maximum denoted by $\eta(p^*)$ if the efficiency function $f(p)$ has a sigmoidal shape.
\end{theory}

{\bf Proof : } Consider the asymptotic cases when $p \to 0$ and $p \to \infty$, we have the limiting cases as $\lim_{p \to 0} f(p) = 0$ and $\lim_{p \to \infty}f(p) = 1$ respectively.
\\ $\bullet$ When $p \to 0$: We have $\lim_{p \to 0}\Phi=1$ trivially and $\lim_{p \to 0}\eta=0$.
\\ $\bullet$ When $p \to \infty$: $\lim_{p \to \infty}\Phi=0$ and $\lim_{p \to \infty}\eta=0$.
\\Thus from the extension of the mean value theorem proposed in \cite{meanvalue}, we have $\frac{\mathrm{d}\eta}{\mathrm{d}p}=0$ for at least one $p$.

Consider the function $\frac{1}{\eta}=A(p) + B(p)$, where $A(p)=\frac{p}{f(p)R}$ and $B(p)=\frac{b}{Rq(1-\Phi)}$. From the earlier work in \cite{rodriguez}, we have that $A(p)$ is convex and that $\frac{1}{A(p)}$ is quasi-convex and has a unique maximum at $p_0^*$. $\frac{\mathrm{d}f(p)}{\mathrm{d}p}>0$  for all $p$. We also know that for $p>p^*_0$, and  $\frac{\mathrm{d}^2 f(p)}{\mathrm{d}p^2}<0$.

Now let us study the derivatives of the function $B(p)$.
\begin{equation}
\frac{\mathrm{d}B(p)}{\mathrm{d}p} = \frac{b}{Rq(1-\Phi)^2}\frac{\mathrm{d}\Phi}{\mathrm{d}p}
\label{eq:d1bp}
\end{equation}
and we have
\begin{equation}
\frac{\mathrm{d}^2B(p)}{\mathrm{d}p^2} = \frac{b}{Rq(1-\Phi)^2}(\frac{\mathrm{d}^2\Phi}{\mathrm{d}p^2} +\left(\frac{\mathrm{d}\Phi}{\mathrm{d}p})^2 \frac{2}{1-\Phi} \right).
\label{eq:d2bp}
\end{equation}

If $B(p)$ is a monotonically decreasing function and is convex for $p \geq p_0^*$, then we have $\frac{1}{A(p)+B(p)}$ to be quasi-concave \cite{boyd}. So in the following section of this proof we will show that $\frac{dB(p)}{dp} <0$ and $\frac{\mathrm{d}^2B(p)}{\mathrm{d}p^2}>0$. \\
For $\frac{\mathrm{d}B(p)}{\mathrm{d}p} <0$, by examining equation \ref{eq:d1bp}, we see that showing $\frac{\mathrm{d}\Phi}{\mathrm{d}p} <0$ is sufficient as the other terms are always positive.\\
Similarly, for  $\frac{\mathrm{d}^2B(p)}{\mathrm{d}p^2} >0$, by examining equation \ref{eq:d2bp}, we see that showing  $\frac{\mathrm{d}^2\Phi}{\mathrm{d}p^2} >0$ is sufficient as the other terms are also always positive.\\
\begin{equation}
\frac{\mathrm{d}\Phi}{\mathrm{d}p} = (1-f(p))\frac{\mathrm{d}\Pi_K}{\mathrm{d}p} - \Pi_K \frac{\mathrm{d}f(p)}{\mathrm{d}p}.
\label{eq:d1pip}
\end{equation}

\begin{equation}
\frac{\mathrm{d}^2\Phi}{\mathrm{d}p^2} = (1-f(p))\frac{\mathrm{d}^2\Pi_K}{\mathrm{d}p^2} -2\frac{\mathrm{d}\Pi_K}{\mathrm{d}p}\frac{\mathrm{d}f(p)}{\mathrm{d}p}-\Pi_K\frac{\mathrm{d}^2f(p)}{\mathrm{d}p^2}.
\label{eq:d2pip}
\end{equation}
For  $\frac{\mathrm{d}\Phi}{\mathrm{d}p} <0$, by examining equation \ref{eq:d1pip}, we see that showing  $\frac{\mathrm{d}\Pi_K}{\mathrm{d}p} <0$ is sufficient.\\
We have $\rho=\frac{q}{1-q}\frac{1-f(p)}{f(p)}$ and $\frac{\mathrm{d}\rho}{\mathrm{d}p}=\frac{-q}{(1-q)f(p)^2}\frac{\mathrm{d}f(p)}{\mathrm{d}p}$ which is negative.
Express $\frac{1}{\Pi_K}= 1+\frac{1}{\rho}+..\frac{1}{\rho^K}$. Differentiating with respect to $p$, we have
\begin{equation}
\frac{\mathrm{d}\Pi_K}{\mathrm{d}p} = \Pi_K^2 \frac{\mathrm{d}\rho}{\mathrm{d}p}(\frac{1}{\rho^2}+..\frac{K}{\rho^{K+1}}).
\label{eq:dpipexp}
\end{equation}

Similarly, for  $\frac{\mathrm{d}^2\Phi}{\mathrm{d}p^2} >0$, by examining equation \ref{eq:d2pip}, we see that showing $\frac{\mathrm{d}^2\Pi_K}{\mathrm{d}p^2} >0$ is sufficient as $\frac{\mathrm{d}^2f(p)}{\mathrm{d}p^2}<0$ for $p>p^*_0$ and $\frac{\mathrm{d}\Pi_K}{\mathrm{d}p} <0$. And from equation \ref{eq:dpipexp} we observe that as $p$ increases $\frac{\mathrm{d}\Pi_K}{\mathrm{d}p} $  increases. Thus following the argument from the start, we have $\eta(p)$ to be quasi-concave.\\ Since there exists some power $p$ for which $\eta(p)$ is maximized, we have proved that there exists a unique $p^*$ for which the energy efficiency is optimized.
\endpf

We are then able to determine the optimal power $p^*$ which maximize the energy efficiency function, by solving the following equation:
\begin{equation}
0 = \frac{-\mathrm{d}\Phi}{\mathrm{d}p}\{b + \frac{pq(1-\Phi)}{f(p)}\} + (1-\Phi)\{ \frac{\mathrm{d}\Phi}{\mathrm{d}p}\frac{p}{f(p)} +\frac{\mathrm{d} (p/f(p) )}{\mathrm{d}p}\}.
\label{eq:etad}
\end{equation}

\subsection{Behavior of $p^*$ with respect to $q$}

When $q \to 0$, the optimization problem is reduced to finding $p$ that maximizes $\frac{f(p)}{p}$. This can be obtained by calculating the derivative of $p^*$ from equation \ref{eq:etad} and applying the limit on $q$. Indeed we have that $\lim_{q \to 0}  \frac{\mathrm{d}\Phi}{\mathrm{d}p}=0$ and then equation (\ref{eq:etad}) is reducing to $\frac{\mathrm{d}(p/f(p) )}{\mathrm{d}p}=0$.

When $q \to 1$, the optimization problem is reduced to finding $p$ that maximizes $\frac{f(p)}{p+b}$. This can be obtained by simply using same ideas as previously.
%Note that if $b_2 > b_1$, then $p^*(q=1,b2)>p^*(q=1,b1)$.

\subsection{Power Control with the QoS and maximum power constraint}
The QoS constraint requires that $\Phi \leq \epsilon$ and then we have to find the new properties of the energy efficiency function satisfying this constraint. We define $p_0:=\min(p|\Phi(p) \leq \epsilon)$.
\begin{propposition}
For all $p>p_0$, the constraint is satisfied, i.e., $\Phi(p>p_0) \leq \epsilon$.
\end{propposition}

{\bf Proof :} This is quite easy to see because from our earlier proof we have that $\frac{\mathrm{d}\Phi}{\mathrm{d}p}<0$ and so $\Phi(p) \leq \Phi(p_0) \leq \epsilon$ for all $p>p_0$.
\endpf

Additionally we also have another proposition which gives the properties of $p_0$ when the arrival probability, $q$, changes:

\begin{propposition}
If $q_2>q_1$, then $p_0(q_2) \geq p_0(q_1)$.
\end{propposition}
{\bf Proof :}  This result can also be easily proved. From our earlier proof we have $\frac{\mathrm{d}\Phi}{\mathrm{d}q}>0$ and so with the power $p_0(q_2)$, we have $\Phi(q_1) \leq \Phi(q_2) \leq \epsilon$. Thus $p_0(q_1) \leq p_0(q_2)$.
\endpf

With these results, we show in the following proposition that the energy efficiency function with the constraint, denoted by $\eta_*$, can still be optimized and has a unique maximum.

\begin{propposition}
Given $\eta(p)$ with a unique  $p^*$ and a constraint on $\Phi$ as $\Phi \leq \epsilon$, satisfied by $p \geq p_0$; the modified energy efficiency $\eta_*$ has a unique  $p^*_* = \min[\max(p_0,p^*),p_{\max}]$.
\end{propposition}
{\bf Proof :} To proceed with our proof we solve the problem using the KKT conditions (see e.g., \cite{boyd}). The Lagrangian is defined by:
\begin{equation}
\mathcal{L}(p, \lambda_1, \lambda_2, \lambda_3) =  -\eta(p) + \lambda_1 (\Phi -L)
+ \lambda_2 (p - P_{\max}) - \lambda_3 p.
\end{equation}
The KKT conditions applied to the above quasi-convex optimization problem yield 
\begin{equation}
\begin{array}{rcl}
\frac{\mathrm{d} \mathcal{L}}{\mathrm{d} p} & =& 0\\
\Phi - L & \leq & 0\\
p - P_{\max} & \leq & 0\\
-p & \leq & 0\\
\lambda_1 (\Phi-L) & = & 0\\
\lambda_2 (p - P_{\max}) & = & 0\\
\lambda_3 (-p) & = 0 \\
\lambda_1 & \geq 0 \\
\lambda_2 & \geq 0 \\
\lambda_3 & \geq 0 \\
\end{array}
\end{equation}

\endpf

\section {Numerical results}

In this section we use simulations and numerical calculations to study the effectiveness of our proposal as well as the advantages offered.

\subsection{Convergence to steady state}

So far in our model, we assume the system to be in the steady state. However, in reality, it takes some time for the system to reach the steady state. In order to study the relationship between the observed values for packet loss, and the theoretical values, we devise the following simulation: Using random number generators and a virtual queue we study the fraction of packets lost for a fixed packet count (representing time) to see how fast the simulated queue converges to the steady state. For each total packet count, the simulation is iterated $10^6$ times for a queue size of $K=10$. We observe that with a packet count of about one thousand, the simulated values of $\Phi$ are on an average the theoretically predicted value ($\pm 4\%$).
%\begin{figure}[H]
%    \begin{center}
%        \includegraphics[width=100mm]{Convergence_comparison.png}
%    \end{center}
%    \caption{ Convergence to the steady state values of $\Pi_K$ or $\Phi$ plotted as $100\frac{\Phi_{Simulation}}{\Phi_{Steady}}$. For a queue size of 10, a packet count of 1000 is required.}
%    \label{fig:simolowt}
%\end{figure}

\subsection{Energy efficiency and power control}

In this section we present some numerical results that bring to focus the advantage of this cross layer approach over a purely physical layer approach.

In the following section we consider the transmitter-receiver pair to be a single input single output link with the success function $f(p)=\exp(-(2^{\frac{R}{R_0}}-1)\frac{\sigma^2}{p})$. We also always consider a queue of maximum size $K=10$, $P_{max}=35$dBm,$P_{min}=10dB$, $R=4000$bps and $R_0=1000$. 
Note that as we have $q=1$ case being identical in theory to the case where we just model the energy efficiency with a purely physical layer approach as in \cite{goodman} (after accounting for the fixed power consumption of the transmitter $b$); if we optimize the energy efficiency it will be optimal to use the power $p^*(q=1)$. However if we consider the cross layer model the energy efficiency is optimized at a different $p^*$ based on the transition probability $q$. As $p^* \leq p^*(q=1)$, using the cross layer optimization we have a gain which can be expressed as $10\log_{10}(\frac{p^*(q=1)}{p^*})$ in dB.

\begin{figure}[H]
    \begin{center}
        \includegraphics[width=100mm]{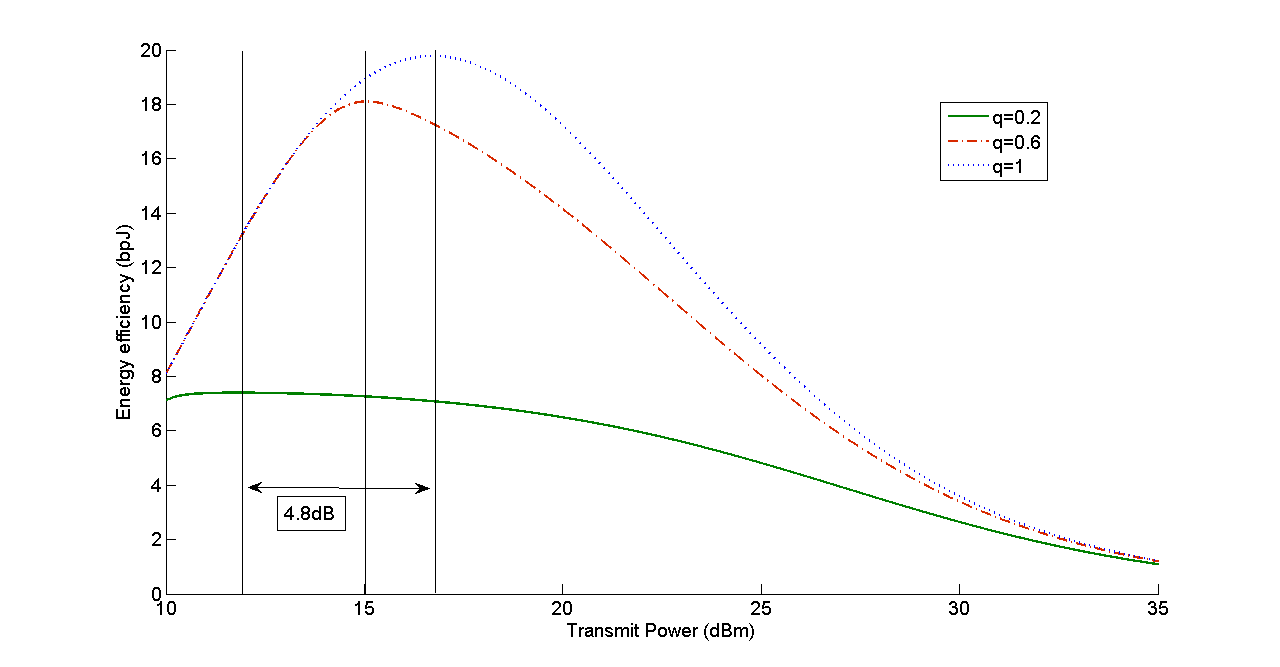}
    \end{center}
    \caption{$\eta$ vs $p$ of a system with $\frac{b}{\sigma^2}=100$ (20dB). Observe that the function is quasi-concave for all $q$ and that $p^*$ decreases as $q$ decreases. }
    \label{fig:eebq}
\end{figure}

In Figure \ref{fig:eebq} we study the energy efficiency of a system with $\frac{b}{\sigma^2}=100$. Here we see that as $q$ decreases $p^*$ decreases. Also seen from the same figure is the quasi-concavity of the energy efficiency function and the asymptotic behavior.

\begin{figure}[H]
    \begin{center}
        \includegraphics[width=100mm]{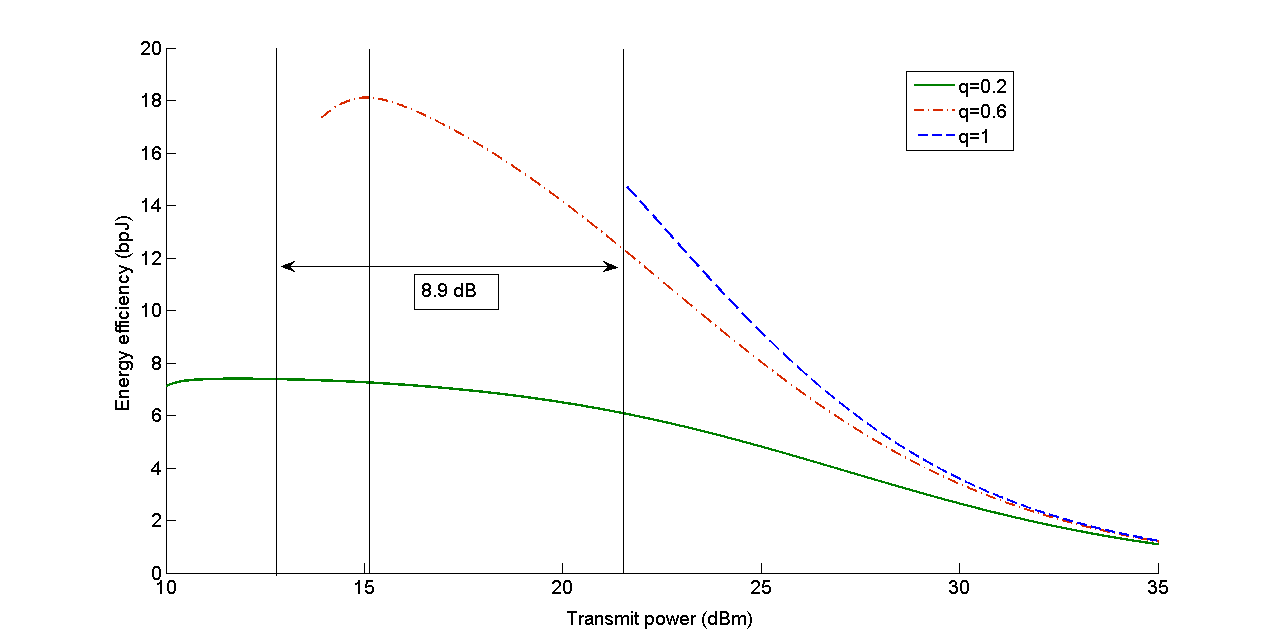}
    \end{center}
    \caption{ EE of system with $\frac{b}{\sigma^2}=100$ (20dB) and $L=0.01$. Note that in this plot, the quasi-concavity is retained and that $p_0$ increases with $q$}
    \label{fig:eet}
\end{figure}

In Figure \ref{fig:eet} we study the energy efficiency of a system with $\frac{b}{\sigma^2}=100$ (20dB) and with a packet loss constraint of $L=0.01$, i.e., $\Phi \leq 0.01$. Here we see that as $q$ decreases the minimum power required to satisfy the constraint. The quasi-concavity of the energy efficiency function is clearly preserved after the constraint and it has a unique maximum.

\begin{figure}[H]
    \begin{center}
        \includegraphics[width=100mm]{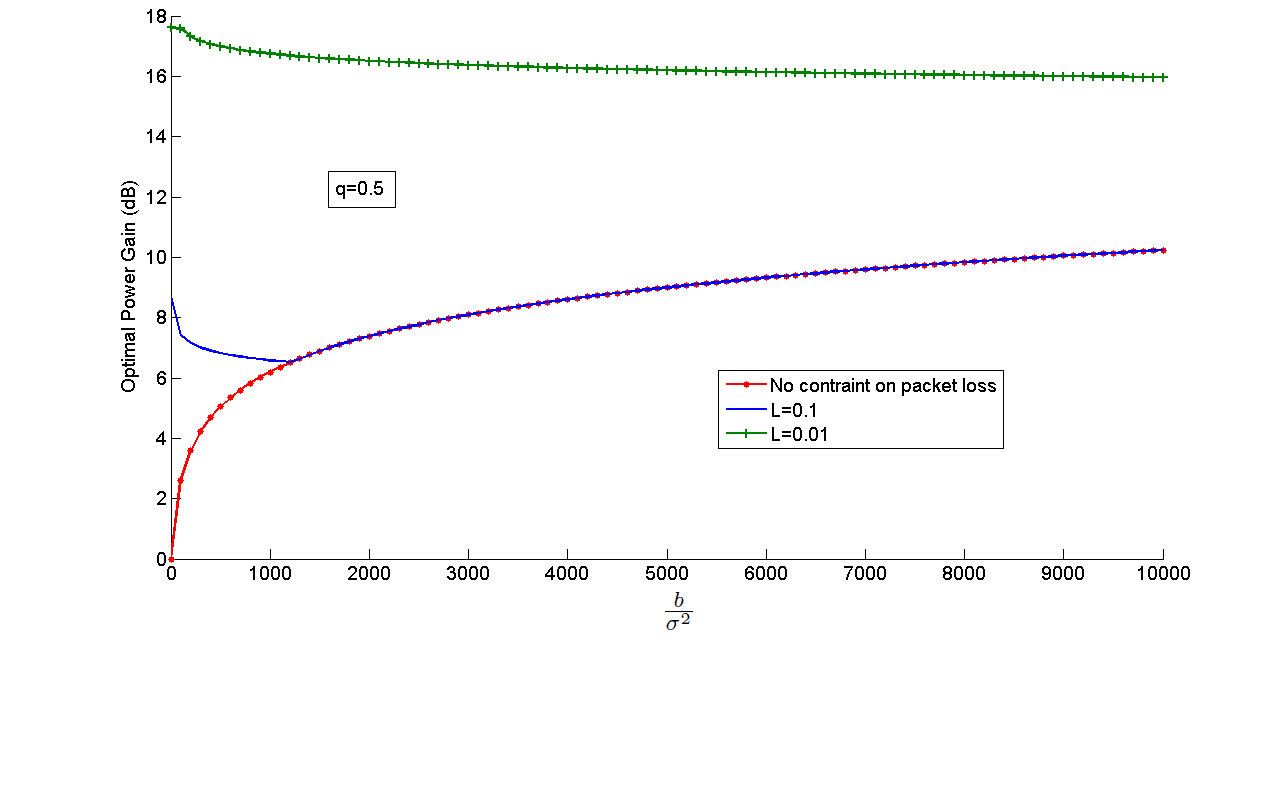}
    \end{center}
    \caption{ The gain in the optimal power while using a cross layer model as opposed to a purely physical layer model i.e $\frac{p^*(q=1)}{p^*}$ plotted against $\frac{b}{\sigma^2}$. }
    \label{fig:gainbs}
\end{figure}

In Figure \ref{fig:gainbs} we study the gain in power with $q=0.5$ plotted against $\frac{b}{\sigma^2}$. For low values of $\frac{b}{\sigma^2}$ we see that the gain for $\epsilon=0.1$ is due to the constraint causing it to decrease with $\frac{b}{\sigma^2}$, however beyond a certain value of $\frac{b}{\sigma^2}$, even the efficiency function for $q=0$ is optimized at $p^*$ (the constraint is met for $p \leq p^*$), the gain is due to the difference in $p^*$ which increases with $\frac{b}{\sigma^2}$ just like for the $\epsilon=1$ case.

\begin{figure}[H]
    \begin{center}
        \includegraphics[width=100mm]{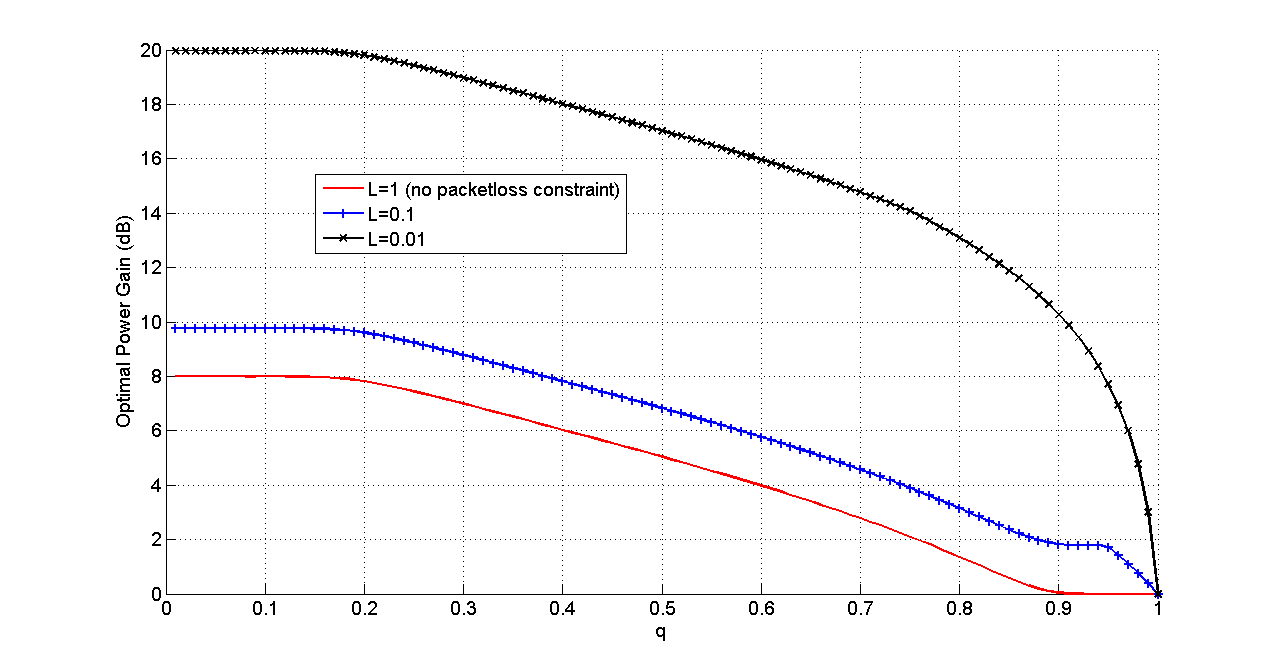}
    \end{center}
    \caption{ The gain in the optimal power while using a cross layer model as opposed to a purely physical layer model i.e $\frac{p^*(q=1)}{p^*}$ plotted against $q$.}
    \label{fig:gainq}
\end{figure}

In Figure \ref{fig:gainq} we study the gain in power with $\frac{b}{\sigma^2}=100$ plotted against $q$. For low values of $q$ we clearly see that the gain in using the cross layer approach is the highest and decreases with $q$.

\subsection{Application of the results on some useful cases}

In a realistic situation, when there is no packet to transmit, a base station consumes about $50\%$ of the power it consumes at full load \cite{salah}. %Thus, $b \approx p_{max}$ if the user receives a  SNR of $30dB$, $\frac{b}{\sigma^2}=30$dB. 
On the other hand, the entry probability $q$ can change based on the service, protocol, traffic, etc:
\begin{itemize}
\item For $q=0.5$, $R=256$Kbps and $R_0=64$Kbps, our numerical calculations show that, for an SNR of 30 dB, using $p^*=3\%$ of the transmit power is optimal. While if the user was at a distance where the received SNR is $20$dB, using $p^*=13\%$ of the power is optimal. The relationship between the optimal powers is clearly not linear with the SNR. 
\item Consider now a system with $q=\frac{1}{25}$ like in some streaming systems that have data sent in one out of 25 frames. In this case, for the user with a SNR of $30$dB, $p^*=1.5\%$ and for $20$dB, $p^*=15\%$. The explanation for this can be seen from the theoretical section, as $q \to 0$, optimizing $\eta$ is the same as optimizing $\frac{f(p)}{p}$ which has a solution that is linear with the SNR.
\end{itemize}

\section{Conclusion}

We have examined the energy efficiency function considering the packet level dynamics of a system and incorporated the effect of the finite buffer size. We find that modeling the system in this form changes the shape of the energy efficiency function. However the energy efficiency function retains its property of quasi-concavity and has a unique maximum. In this work, we also observe that if the packet entry probability is small, the energy efficiency is deformed to a higher extent causing the optimal power to be smaller than in a model ignoring the packet level dynamics. This deformation is due to the constant power consumption of the transmitter even when it does not transmit. The effect of the constant power consumption decreases as the path loss or noise increases and in fact, it is the ratio between the constant power consumption and the noise (with path loss) that determines the deformation. If we impose a constraint on the packet loss, clearly the buffer helps in decreasing this loss which causes a further deformation in the shape of the energy efficiency function.

Many extensions of the proposed work are possible. The most relevant extension is to apply the proposed framework to the case of distributed power control over multiple access channels; we already know that the existence of a pure Nash equilibrium is guaranteed due to quasi-concavity of the energy-efficiency \cite{nashequn}. Another natural extension of the proposed framework is of course, the problem of resource allocation, which is known to be non-trivial, the problem of power allocation is indeed important in multi-carrier and MIMO systems. It would also be fully relevant to study distributed dynamics of the queues and power control policies leading to steady states of the system. Another interesting aspect concerns the case of variable transmission rate as a fixed transmission rate is indeed not the best scheme to minimize energy consumption. 

\section{Acknowledgements}

This work is a joint collaboration between Laboratoire des signaux et syst\'{e}mes (L2S) of Sup\'{e}lec, the Alcatel Lucent Chair of Sup\'{e}lec, Orange Labs R$\&$D as well as University of Avignon. This work is also partially supported by ``L'Agence Nationale de la Recherche" (ANR) via the program Inter Carnot within the project ANR-09-VERS0: ECOSCELLS.


\begin{thebibliography}{9999}

\bibitem{salah} L. Saker and S-E. Elayoubi, ``Sleep mode implementation issues in green base stations", IEEE PIMRC 2010, Istanbul, September 2010

\bibitem{Kim00}J. Kim, M. Honig, ``Resource Allocation for Multiple Classes of DS-CDMA Traffic", in Transactions on Vehicular Technology, vol. 49, no. 2, 2000.

\bibitem{Mesh09}F. Meshkati, H. Poor, S. Schwartz, ``Energy Efficiency-Delay Tradeoffs in CDMA Networks: A Game-Theoretic Approach", in Transactions on Information Theory, vol. 55, no. 7, 2009.

\bibitem {goodman}D. J. Goodman, and N. Mandayam, ``Power Control for Wireless Data", IEEE Personal Communications, vol. 7, pp. 48-54, Apr. 2000.

\bibitem{veronica2} E.V. Belmega and S. Lasaulce, ``Energy-efficient precoding for multiple-antenna terminals", IEEE. Trans. on Signal Processing, 59, 1, Jan. 2011.

\bibitem{valuetools}G. Bacci, A. Bulzomato, M. Luise, ``Uplink power control and subcarrier assignment for an OFDMA multicellular network based on game theory", in Proc. Int. Conf. on Performance Evaluation Methodologies and Tools (ValueTools), Paris, France, May 2011.


\bibitem{honig} M. L. Honig, ``Adaptive linear interference suppression for packet DS-CDMA," European Trans. Telecommun., vol. 9, no. 2, pp. 173-181,
Mar.-Apr. 1998.

\bibitem{kumar} A. Sampath, P. S. Kumar, and J. M. Holtzman, ``Power control and resource management for a multimedia CDMA wireless system," in Proc.
IEEE PIMRC, vol. 1, Toronto, Canada, Sept. 1995, pp. 21-25.

\bibitem{yao} S. Yao and E. Geraniotis, ``Optimal power control law for multi-media multi-rate CDMA systems," in Proc. IEEE VTC, vol. 1, Atlanta, GA,
Apr. 1996, pp. 392-396.

\bibitem{shen} Q. Shen and W. A. Krzymien, ``Power assignment in CDMA personal
communication systems with integrated voice/data traffic," in Proc.
IEEE GLOBECOM, London, U.K., Nov. 1996, pp. 168-172.

\bibitem{rodriguez} V. Rodriguez, ``An Analytical Foundation for Ressource Management in Wireless Communication", IEEE Proc. of Globecom, San
Francisco, CA, USA, pp. 898-902, , Dec. 2003.

\bibitem{sampath} A. Sampath, N. B. Mandayam, and J. M. Holtzman, ``Analysis of an access control mechanism for data traffic in an integrated voice/data wireless CDMA system," in Proc. IEEE VTC, vol. 3, Atlanta, GA, Apr. 1996,
pp. 1448-1452.

\bibitem{wolff} R. W. Wolff, Stochastic Modeling and the Theory of
Queues. Englewood Cliffs, NJ: Prentice-Hall, 1989.

\bibitem{richter} F. Richter, A. J. Fehske, G. Fettweis, ``Energy E 
ciency Aspects of BaseStation Deployment Strategies for Cellular Networks", Proceedings of VTC Fall'2009

\bibitem{boyd} S. Boyd, and L. Vandenberghe, ``Convex Optimization", Cambridge University Press, 2004.

\bibitem{meshkatiCDMA06} F. Meshkati, M. Chiang, H. Poor and S. Schwartz, ``A game-theoretic approach to energy-efficient power control in multicarrier CDMA systems ", in JSAC, vol. 24, no. 6, 2006.

\bibitem{meanvalue}W. Eric, ``Mean-Value Theorem". MathWorld. Wolfram Research.

\bibitem{Kim00} J. Kim and M. Hoenig, ``Resource Allocation for Multiple Classes of DS-CDMA Traffic" in IEEE Transactions on vehicular technology, vol. 49, no. 2, 2000.

\bibitem{AltmanCons07} E. Altman , K. Avrachenkov, N. Bonneau,
M. Debbah, R. El-Azouzi, D. Menasche, ``Constrained Stochastic Games in Wireless Networks", in proceedings of GlobeCom, 2007.

\bibitem{nashequn} J. Rosen, ``Existence and uniqueness of equilibrium points for concave n-person games," Econometrica, vol. 33, pp. 520-534, 1965.

\end{thebibliography}
\end{document}